\documentclass[preprint]{aastex631}

\renewcommand{\baselinestretch}{1.1}
\usepackage{textcomp}
\usepackage{gensymb}
\usepackage{epstopdf}
\usepackage{amsmath}
\usepackage{graphicx}
\usepackage{hyperref}
\usepackage{multirow}
\usepackage[nameinlink]{cleveref}
\usepackage[multiple]{footmisc}
\crefname{paragraph}{\S}{\S\S} 
\shorttitle{MOA-2011-BLG-262} 
\shortauthors{Terry et al.}

\begin{document}

\title{\textbf{A Candidate High-Velocity Exoplanet System in the Galactic Bulge}}

\author[0000-0002-5029-3257]{Sean K. Terry}
\affiliation{Department of Astronomy, University of Maryland, College Park, MD 20742, USA}
\affiliation{Code 667, NASA Goddard Space Flight Center, Greenbelt, MD 20771, USA}

\author[0000-0003-0014-3354]{Jean-Philippe Beaulieu}
\affiliation{Sorbonne Universit\'e, CNRS, Institut d’Astrophysique de Paris, IAP, F-75014 Paris, France}
\affiliation{School of Natural Sciences, University of Tasmania, Private Bag 37 Hobart, Tasmania, 7001, Australia}

\author[0000-0001-8043-8413]{David P. Bennett}
\affiliation{Department of Astronomy, University of Maryland, College Park, MD 20742, USA}
\affiliation{Code 667, NASA Goddard Space Flight Center, Greenbelt, MD 20771, USA}

\author{Aparna Bhattacharya}
\affiliation{Department of Astronomy, University of Maryland, College Park, MD 20742, USA}
\affiliation{Code 667, NASA Goddard Space Flight Center, Greenbelt, MD 20771, USA}

\author[0000-0002-7423-8615]{Jon Hulberg}
\affiliation{Department of Physics, Catholic University of America, Washington, DC 20064, USA}

\author[0000-0003-4591-3201]{Macy J. Huston}
\affiliation{Department of Astronomy, University of California Berkeley, Berkeley, CA 94720, USA}

\author[0000-0003-2302-9562]{Naoki Koshimoto}
\affiliation{Department of Earth and Space Science, Graduate School of Science, Osaka University, Osaka, 560-0043, Japan}

\author[0000-0001-5860-1157]{Joshua W. Blackman}
\affiliation{Physikalisches Institut, Universit{\"a}t Bern, Gesellschaftsstrasse 6, CH-3012 Bern, Switzerland}

\author[0000-0002-8131-8891]{Ian A. Bond}
\affiliation{School of Mathematical and Computational Sciences, Massey University, Auckland 0632, New Zealand}

\author[0000-0003-0303-3855]{Andrew A. Cole}
\affiliation{School of Natural Sciences, University of Tasmania, Private Bag 37 Hobart, Tasmania, 7001, Australia}

\author[0000-0001-9611-0009]{Jessica R. Lu}
\affiliation{Department of Astronomy, University of California Berkeley, Berkeley, CA 94720, USA}

\author[0000-0003-2388-4534]{Cl\'ement Ranc}
\affiliation{Sorbonne Universit\'e, CNRS, Institut d’Astrophysique de Paris, IAP, F-75014 Paris, France}

\author[0000-0002-1530-4870]{Natalia E. Rektsini}
\affiliation{Sorbonne Universit\'e, CNRS, Institut d’Astrophysique de Paris, IAP, F-75014 Paris, France}
\affiliation{School of Natural Sciences, University of Tasmania, Private Bag 37 Hobart, Tasmania, 7001, Australia}

\author[0000-0002-9881-4760]{Aikaterini Vandorou}
\affiliation{Department of Astronomy, University of Maryland, College Park, MD 20742, USA}
\affiliation{Code 667, NASA Goddard Space Flight Center, Greenbelt, MD 20771, USA}

\correspondingauthor{S. K. Terry}
\email{skterry@umd.edu}

\begin{abstract}

\small \noindent We present an analysis of adaptive optics (AO) images from the Keck-I telescope of the microlensing event MOA-2011-BLG-262. The original discovery paper by \cite{bennett:2014a} reports two distinct possibilities for the lens system; a nearby gas giant lens with an exomoon companion or a very low mass star with a planetary companion in the galactic bulge. The $\sim$10 year baseline between the microlensing event and the Keck follow-up observations allows us to detect the faint candidate lens host (star) at $K = 22.3$ mag and confirm the distant lens system interpretation. The combination of the host star brightness and light curve parameters yields host star and planet masses of $M_{\rm host} = 0.19 \pm 0.03M_{\sun}$ and $m_p = 28.92 \pm 4.75M_{\Earth}$ at a distance of $D_L = 7.49 \pm 0.91\,$kpc. We perform a multi-epoch cross reference to \textit{Gaia} DR3 and measure a transverse velocity for the candidate lens system of $v_L = 541.31 \pm 65.75$ km s$^{-1}$. We conclude this event consists of the highest velocity exoplanet system detected to date, and also the lowest mass microlensing host star with a confirmed mass measurement. The high-velocity nature of the lens system can be definitively confirmed with an additional epoch of high-resolution imaging at any time now. The methods outlined in this work demonstrate that the \textit{Roman} Galactic Exoplanet Survey (RGES) will be able to securely measure low-mass host stars in the bulge.
\\
\\
\textit{Subject headings}: gravitational lensing: micro, planetary systems \\
\end{abstract}


\section{Introduction} \label{sec:intro}
\indent Of the more than 5600 extrasolar planets discovered to date, most have relatively short orbital periods because the primary detection methods (e.g. transits, radial velocity) are most sensitive to these close-in planets. On the other hand, the gravitational microlensing method is most sensitive to longer period planets that orbit beyond the snow line \citep{mao:1991a, gould:1992a}. The core accretion theory \citep{lissauer93, pollack:1996aa} predicts that the most massive planets in systems should form beyond this snow line. The theory states that a runaway gas accretion phase happens once a planetary core has accumulated ${\sim}10 M_{\oplus}$ of solid rocky material. The planet will then rapidly accumulate nearby Hydrogen and Helium gas onto its surface until the surrounding gas has been depleted. This runaway accretion phase results in gas giant planets with masses between Saturn (${\sim}\, 90M_{\oplus}$) and Jupiter (${\sim}\, $320$M_{\oplus}$). As a result of this, some leading theories of planet formation \citep{ida:2004a, mordasini:2009a} predict a dearth of intermediate-mass giant planets ($20 - 80M_{\oplus}$) at wide orbits. \\
\indent One of the largest statistical studies of the microlensing exoplanet population, performed by \cite{suzuki:2016a, suzuki:2018a}, found a gap in the planet-star mass ratio distribution between $1\times10^{-4} < q < 4\times10^{-4}$. We underline the \cite{suzuki:2016a, suzuki:2018a} studies measured the mass-ratio distribution and not the true planet mass distribution, however recent work by \cite{bennett:2021a} reports evidence for a lack of this sub-Saturn planet mass desert by analyzing a sample of radial velocity exoplanets from the CORALIE and HARPS projects \citep{mayor:2011a}. Recent studies by \cite{adams:2021a} and \cite{ali:2022a} cast further doubt on the sub-Saturn desert scenario by studying the effects of slowed accretion and angular momentum loss by giant planet formation in simulations, as well as the effect of late giant collisions on the subsequent formation of cold sub-Saturns. Lastly, we note upcoming work by the KMTNet \citep{kim:2016a} group which claims a gap in the mass-ratio distribution at approximately $-3.6 < log(q) < -3$, corresponding to Saturn$-$Jupiter masses \citep{gui:inprep, zang:inprep}. This new claimed gap is not predicted by any core accretion theory of planet formation at present, and there remain open questions regarding the differences in detection efficiency between KMTNet planets in this mass range from their 2016-2019 sample and 2021 sample (e.g. same number of mass-gap planets in the four year dataset and one year dataset). One final distinction with these mass-ratio gap claims is that nearly all microlensing discovery papers implement a Bayesian analysis that relies on the assumption that stars of any mass are equally likely to host a planet with the given mass ratio. \\
\indent The work presented here is part of the NASA Keck Key Strategic Mission Support (KSMS) program, ``Development of the WFIRST Exoplanet Mass Measurement Method" \citep{bennett_KSMS}, which is a pathfinder project for the \textit{Nancy Grace Roman Space Telescope}  (formerly known as \textit{WFIRST}) \citep{spergel:2015a}. A large fraction of the \textit{Roman} Telescope observing time will be devoted to the \textit{Roman} Galactic Bulge Time Domain Survey (\textit{GBTDS}) \citep{Gaudi:2022a}. As part of this bulge survey, the Roman Galactic Exoplanet Survey (\textit{RGES}) will be the first dedicated space-based gravitational microlensing survey and is expected to detect over 30,000 microlensing events and over 1,400 bound exoplanets during its five-year survey \citep{penny19}. The survey is also expected to discover several hundred free-floating planets (FFPs) and/or planets on very wide orbits around host stars. The primary goal of the KSMS program is to determine masses and distances for a large majority of stars in the \cite{suzuki:2016a} statistical sample. Many of these successful measurements have now been published \citep{bennett:2015a, batista:2015a, bennett:2020a, bhattacharya:2021a, terry:2021a, terry:2022a}, with more confirmed mass measurements in preparation \citep{bhattacharya:inprep, vandorou:inprep}. Several events have been found to have host masses at the $>$90th percentile of the Bayesian prediction based on the equal planet hosting probability assumption mentioned earlier. These are events MOA-2013-BLG-220 \citep{vandorou:2020a}, MOA-2007-BLG-400 \citep{bhattacharya:2021a}, MOA-2007-BLG-192 \citep{terry:2024a}, OGLE-2016-BLG-1195 \citep{vandorou:2023a}, and OGLE-2012-BLG-0563 \citep{bennett:inprep, bhattacharya:inprep}. This suggests that higher mass stars could be much more likely to host planets beyond the snow line with mass ratios between $0.002 < q < 0.004$. Lastly, the core accretion theory was primarily developed with solar type host stars in mind, the gap expected from the runaway gas accretion scenario might exist for solar-type stars, but be washed out with the low-mass M-dwarf hosts that are ubiquitous in the microlensing sample. Successful mass measurements like the one presented in this paper can continue to probe this possibility.

\subsection{High Velocity \& Hypervelocity Stars}\label{subsec:HVSs}
Historically, studies of high velocity stars (sometimes called `runaway stars') were primarily focused on milky way (MW) stellar halo stars \citep{eggen:1962a}, which have space velocities of ${>}100$ km s$^{-1}$ \citep{keenan:1953a}. Early ideas to explain the formation of these high velocity stars include supernovae in binary star systems \citep{blaauw:1961a} or ejections caused by encounters with star clusters \citep{poveda:1967a}. In the late 1980's, \cite{hills:1988a} proposed a new formation channel that could generate stars with velocities in excess of $1,000$ km s$^{-1}$. These ``hypervelocity stars" (HVS) would originate as typical stellar binary systems that encountered the supermassive black hole (SMBH) at the center of most galaxies. This interaction with the SMBH would capture one of the stars in the binary system, and eject the second star at very high velocity away from the central region of the galaxy. This scenario would become known as the ``Hills Mechanism". Since then, several dozen HVSs in our Galaxy have been detected, however these stars could not be definitively determined to originate from an encounter with SgrA$^*$ and thus are unable to confirm the Hills Mechanism. A few years ago, \cite{koposov:2020a} reported a detection of the highest velocity HVS ($> 1,700$ km s$^{-1}$) and were able to unambiguously trace the star to the MW's galactic center, thus confirming the Hills Mechanism.\\
\indent To date, no exoplanet has been discovered orbiting an HVS. Prior studies have examined the likelihood of an HVS hosting a planet, mostly from a theoretical modeling standpoint. \cite{ginsburg:2012a} performed simulations of HVSs with bound planets generated through the Hills Mechanism, and found that in order for a planet to remain bound to the HVS the initial binary separation would need to be in the range $a_* = 0.05 - 0.5$ AU. Further, the planetary separation would need to be in the range $a_p = 0.02 - 0.05$ AU (e.g. a hot Jupiter). This suggests the planetary systems must be compact in order to survive the strong gravitational encounters. The study of \cite{fragione:2017a} computed the likelihoods of detecting exoplanets around runaway stars and HVSs, and found that the probability of detection depends on many factors including the range of semi-major axes, mean planetary inclination and eccentricities, which is no surprise as these are some of the primary factors governing transit and/or radial velocity detectability in general. The authors use the total number of expected HVS detections from \textit{Gaia} (${\sim}100$; \cite{kenyon:2014a, debruijne2015a}) and the assumption that each HVS hosts a compact planetary system to predict at least one observable transit in this sample. Lastly, it is worth noting these studies considered only the transit and radial velocity methods for detecting the exoplanets in their samples.
\\
\indent In this paper, we present a candidate for the highest velocity planetary system which is also the lowest mass planetary host star residing in the galactic bulge. We successfully measure flux coming directly from the $0.19M_{\odot}$ candidate, which also allows us to obtain the mass of the planetary companion of ${\sim}\, 29M_{\oplus}$. Section \ref{sec:event} describes the original observations for MOA-2011-BLG-262. In Section \ref{sec:follow-up}, we describe the Keck adaptive optics (AO) follow-up analysis that confirms one of the two possible solutions presented in \cite{bennett:2014a}. Section \ref{sec:prop-motion} details our lens-source relative proper motion analysis, as well as our cross-calibration to \textit{Gaia} DR3 to obtain the absolute proper motions for the lens and source. In Sections \ref{sec:light-curve} and \ref{sec:lens-properties} we perform updated modeling of the light curve data and present lens system properties with new constraints from high-resolution imaging. Finally, we discuss the results and conclude the paper in Section \ref{sec:conclusion}.


\section{Microlensing Event MOA-2011-BLG-262} \label{sec:event}
MOA-2011-BLG-262 (hereafter MB11262), located at RA (J2000) $=$ 18:00:23.48, DEC (J2000) $= -$31:14:42.9 and Galactic coordinates ($l,b=(-0.369\degree, -3.924\degree)$), was first alerted by the Microlensing Observations in Astrophysics (MOA; \citealt{bond01,sumi:2003a}) collaboration on 26th June 2011. The short-duration and high-magnification nature of the event led several follow-up networks (PLANET, $\mu$FUN) to observe the target at high cadence primarily in the $V$, $I$, and $H-$bands. \\
\indent The MOA photometric reduction methods have improved since the original discovery paper of \citet{bennett:2014a}, so we have re-reduced the MOA $V$ and $R-$ band photometry. We have used the method of \citet{bond01,bond17} to reduce the data from the MOA-II telescope, the Mt.~John Observatory Boller and Chivens 0.61m telescope (operated by the MOA group). The MOA-II data were corrected for systematic errors due to chromatic differential refraction \citep{bennett12}. New reductions are also needed to provide a Markov Chain Monte Carlo (MCMC) distribution to understand the distribution of light curve models that are consistent with the data. We further describe the updated modeling of this light curve data in Section \ref{sec:light-curve}.\\
\indent \cite{bennett:2014a} also obtained high angular resolution adaptive optics (AO) data from the NIRC2 instrument on the Keck-II telescope, taken on 2012 May 14. In particular, the Keck $H-$band measurement of the source was compared to the CTIO $H-$band measurement and the authors found no significant detection of flux from the lens. Ultimately, \cite{bennett:2014a} report two possible solutions for the MB11262 lens system; either a ${\sim}\, 3M_{J}$ host with a ${\sim}\, 0.5M_{\oplus}$ companion located ${\sim}$500pc from Earth, or a low-mass host star ($0.12M_{\odot}$) with a ${\sim}\, 20M_{\oplus}$ companion residing in the Galactic bulge at $D_L > 7$ kpc. Regarding the ``slow" vs. ``fast" solutions previously reported, the fast solution is possible because the angular source radius matches the separation between the caustic entrance and exit (e.g. Figure 2 of \cite{bennett:2014a}). This merges the normal caustic entry and exit peaks into a single bump that resembles the bump from the case of a source radius much larger than the caustic entry and exit separation (e.g. Figure 3 of \cite{bennett:2014a}).\\
\indent Although a low-mass host star in the bulge was initially disfavored by $\Delta\chi^2\, {\sim}\, 2.9$, the full Bayesian analysis did not favor the exoplanet$+$exomoon interpretation. The preference for a nearby planetary-mass lens was not enough to overcome the preference for a stellar-mass lens because of their larger Einstein radii. When the appropriate mass-function prior was included in the \cite{bennett:2014a} study, a stellar-mass host was preferred. As described in Sections \ref{sec:prop-motion} and \ref{sec:lens-properties} of this work, we are able to definitively rule out the nearby planetary-mass host solution.


\section{Keck Follow-up and Analysis} \label{sec:follow-up}
The target MB11262 was observed with the OSIRIS instrument on Keck-I in the K$_{prime}$ passband ($\lambda_c = 2.12\, \mu$m, hereafter Kp) on May 19, 2021. The Keck data have a point-spread function (PSF) full-width half-max (FWHM) of 58 mas and a Strehl ratio (SR) of 0.37. These two metrics are generally used to understand the performance of the AO system, with smaller FWHM and larger SR values typically indicating better quality. We note there are other factors that can affect AO image quality that are not necessarily captured by FWHM or SR \citep{terry:2023a}. Some of these include static or quasi-static aberrations in the optical path, unaccounted for variations in the atmospheric profile information primarily from ground layer turbulence, brightness variations of the tip/tilt guide star (TTGS) due to high cirrus clouds, dome and secondary structure temperature variations \citep{ramey:2022a} among others. \\
\indent For the 2021 observations, the OSIRIS imager was used, which has a $20\arcsec \times 20\arcsec$ field of view and a pixel scale of 9.952 mas/pixel. All of the images were taken using the Keck-I laser guide star adaptive optics (LGSAO) system. As mentioned previously, \cite{bennett:2014a} obtained Keck-II/NIRC2 images of the target in 2012. While we don't perform a full re-analysis of the 2012 data in this work, we do adopt the NIRC2 medium camera data taken in 2012 to perform the photometric calibration (this Section) and multi-epoch proper motion analysis (Section \ref{sec:gaia_pms}). A co-add of 20 dithered medium camera images were used for photometric calibration to images from the Vista Variables in the Via Lactea (VVV) survey \citep{minniti:2010a} following the procedure of \cite{beaulieu:2018a}. This calibration analysis results in uncertainties of 0.07 magnitudes.

\begin{figure*}[!htb]
\includegraphics[width=1.0\linewidth]{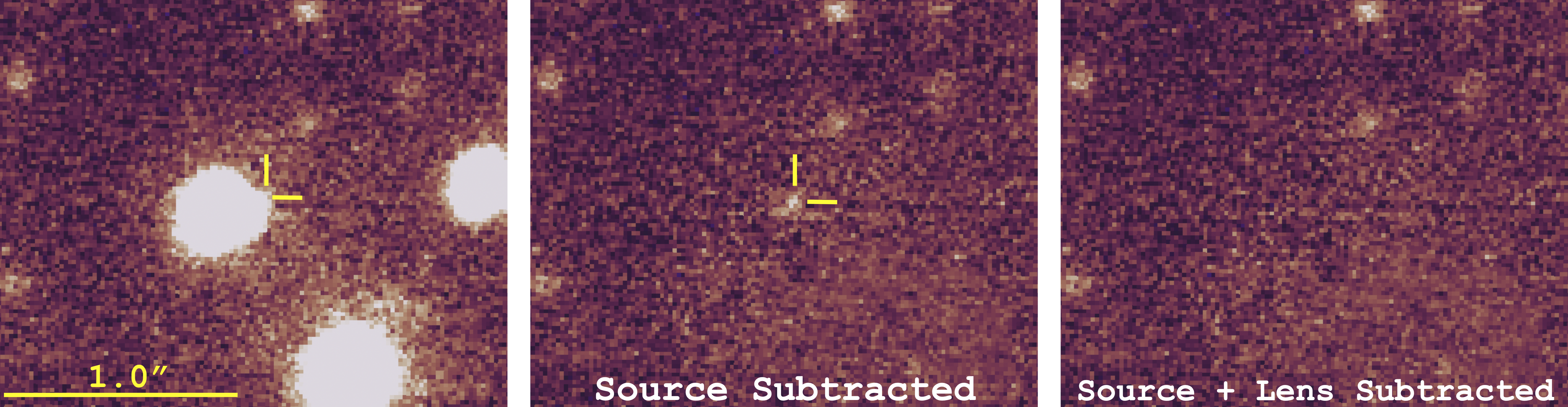}
\centering
\caption{\footnotesize \textit{Left}: A zoom-in on the stacked Kp-band image of MOA-2011-BLG-262, with the brighter source star near center and the faint lens star candidate indicated with the cross-hair. \textit{Middle}: Zoomed-in frame with a single-PSF model fit to the source star and subtracted from the frame, the remaining flux comes primarily from the lens and is indicated with the cross-hair. The other bright nearby stars are also subtracted with the same PSF model. \textit{Right}: Zoomed-in frame with the PSF model fit and subtracted from the source and lens stars. North is up, East is left in all frames. Note there are several faint ambient field stars that are at or below the default detection threshold in \texttt{DAOPHOT}. \label{fig:3panel_keck}}
\end{figure*}

\indent We combined 20 flat-field frames, 10 dark frames, and 8 sky frames for calibrating our science images. A total of 25 science frames with an integration time of 57 s frame$^{-1}$ were reduced using the Keck AO Imaging (KAI) Data Reduction Pipeline \citep{lu:2022a} to correct instrumental aberrations and geometric distortions \citep{lu:2008a, yelda:2010a, freeman:2023a}. Additionally, an independent image-level analysis following the methods of \cite{beaulieu:2018a} was carried out on this dataset, which gives similar results with marginally lower S/N. The stacked science frame, showing a zoomed view of the target can be seen in the left panel of Figure \ref{fig:3panel_keck}. \\

\subsection{PSF Fitting Photometry \& Astrometry}\label{subsec:photometry_astrometry}
Given the large flux ratio and the expected separation between the source and lens stars, we find it necessary to use a PSF fitting routine to measure both targets independently. Following the methods of \cite{bhattacharya:2018a} and \cite{terry:2021a}, we use a modified version of the \texttt{DAOPHOT-II} software package \citep{stetson:1987a} to generate and fit a PSF to the source+lens blend. This routine takes the empirical PSF model and performs Markov Chain Monte Carlo (MCMC) sampling on a pixel grid that encompasses both the blended targets. Additional details of the MCMC routine are given in \citep{terry:2021a, terry:2022a}.\\
\indent The first pass of \texttt{daophot-mcmc} fits a single PSF to the source star and reveals the faint signature of the lens star in the residual image which can be seen in the middle panel of figure \ref{fig:3panel_keck} (labeled ``\texttt{Source Subtracted}"). The approximate position of the signal gives a lens-source separation that is consistent with the ``slow" solution reported by \cite{bennett:2014a}. \\
\indent Re-running \texttt{daophot-mcmc} in the two-star fitting mode produces a smoother residual, shown in the right panel of figure \ref{fig:3panel_keck} (labeled ``\texttt{Source + Lens Subtracted}"). The $\chi^2$ difference for the two-star fit compared to the single-star fit is $\Delta\chi^2 \sim 17$, which is a modest improvement considering the lens star contributes less than 2\% of the total source+lens flux. Table \ref{table:dual-phot} shows the calibrated magnitudes for the two stars of $K_{S}=18.10 \pm 0.07$ and $K_{L}=22.34 \pm 0.15$. Further, Table \ref{table:epoch_seps} reports the measured lens-source separation and relative proper motion in the East and North directions, as well as the lens-source flux ratio. The uncertainties are derived from the ``jackknife" method \citep{quenouille56, tierney1999a}, details of which can be found in \cite{bhattacharya:2021a}, \cite{bennett:2021a}, and \cite{terry:2021a}. Using the near-IR extinction map of \cite{Surot2020} and the \cite{nishiyama:2009a} absorption law, we find a $K$ band extinction of $A_K = 0.169 \pm 0.044$. From our re-analysis of the light curve modeling (Section \ref{sec:light-curve}), we find a source color of $V_S - I_S = 1.90 \pm 0.08$, which leads to an extinction-corrected color of $V_{S0} - I_{S0} = 0.73 \pm 0.08$. We use the color-color relations of \cite{kenyon:1995a} and the I-band magnitude, $I_S = 19.95 \pm 0.09$ to predict a source $K$ band magnitude of $K_S = 17.91 \pm 0.08$. The source magnitude measured in Keck (Table \ref{table:dual-phot}) is slightly fainter than this prediction by ${\sim}\, 1 \sigma$, thus we conclude that there is no evidence of additional flux from a companion to the source.

\begin{deluxetable}{lcr}[!h]
\deluxetablecaption{Two-Star PSF Photometry\label{table:dual-phot}}
\tablecolumns{3}
\setlength{\tabcolsep}{14.0pt}
\tablewidth{\linewidth}
\tablehead{
\colhead{\hspace{-15mm}Star} &
\colhead{Passband} & \colhead{\hspace{19mm}Magnitude}
}
\startdata
Lens & Keck $K$ & $22.34 \pm 0.15$\\
Source & Keck $K$ & $18.10 \pm 0.07$\\
Source $+$ Lens & Keck $K$ & $18.08 \pm 0.06$\\
\enddata
\tablenotetext{}{\footnotesize{\textbf{Note}. Magnitudes are calibrated to the VVV photometric system as described in section \ref{sec:follow-up}.}}
\end{deluxetable}


\section{Lens-Source Relative Proper Motion} \label{sec:prop-motion}
The 2021 Keck-I follow up observations were taken 9.90 years after peak magnification in 2011. The motion of the lens and source on the sky frame is the primary cause for their apparent separation, however there is also a small component that can be attributed to the orbital motion of Earth. As this effect is smaller than $\leq0.1$ mas for a lens at a distance of $D_{L} \geq 7$ kpc, we are safe to ignore this contribution in our analysis as it is much smaller than the error bars on the stellar position measurements. Table \ref{table:epoch_seps} gives the lens-source relative proper motion, which is measured to be $\mu_{\textrm{rel},H} = (\mu_{\textrm{rel,H,E}},\mu_{\textrm{rel,H,N}}) = (-10.926 \pm 0.584, 2.019 \pm 0.321)$ mas yr$^{-1}$. The subscript `H' indicates that these measurements were made in the Heliocentric reference frame, and the `E' and `N' subscripts represent the East and North directions respectively. We also give in Table \ref{table:epoch_seps} the measured lens/source flux ratio of $0.020 \pm 0.003$. This is the largest contrast between a lens/source pair that has been directly measured to-date.

\begin{deluxetable*}{rcccc}[!htb]
\deluxetablecaption{Lens-Source Separation, Relative Proper Motion, and Flux Ratio\label{table:epoch_seps}}
\tablecolumns{3}
\setlength{\tabcolsep}{14.0pt}
\tablewidth{\linewidth}
\tablehead{
\colhead{}&
  \multicolumn{3}{c}{Separation (mas)} &\colhead{Flux Ratio} \\
\cline{2-4}
\colhead{Year} &
\colhead{East} & \colhead{North} & \colhead{Total} & \colhead{[lens/source]}
}
\startdata
2021 & $-108.21 \pm 5.74$ & $19.99 \pm 3.17$ & $110.04 \pm 6.56$ & $0.020 \pm 0.003$\\
\hline
\hline
 & $\mu_{\textrm{rel,H,E}}$(mas/yr) & $\mu_{\textrm{rel,H,N}}$(mas/yr) & $\mu_{\textrm{rel,H}}$(mas/yr)\\
 \hline
 & $-10.93 \pm 0.58$ & $2.02 \pm 0.32$ & $11.12 \pm 0.66$\\
\enddata
\centering \tablenotetext{}{\footnotesize{\textbf{Note}. Errors are derived from the ``jackknife" method as described in section \ref{subsec:photometry_astrometry}.}}
\end{deluxetable*}

\indent Light curve modeling (section \ref{sec:light-curve}) is most conveniently performed in the Geocentric reference frame that moves with the Earth at the time of the event peak. Thus, we must convert between the Geocentric and Heliocentric frames by using the relation given by \cite{dong:2009b}:

\begin{equation}\label{eq:mu-rel}
\mu_{\textrm{rel,H}} = \mu_{\textrm{rel,G}} + \frac{{\nu_{\Earth}}{\pi_{\textrm{rel}}}}{AU} \ ,
\end{equation}

\noindent where $\nu_{\Earth}$ is Earth's projected velocity relative to the Sun at the time of peak magnification. For MB11262 this value is $\nu_{\Earth \textrm{E,N}} = (29.187, -0.438)$ km/sec = $(6.153, -0.092)$ AU yr$^{-1}$ at HJD$' = 5739.13$. With this information and the relative parallax relation $\pi_{\rm{rel}} \equiv 1/D_{L} - 1/D_{S}$, we can rewrite equation \ref{eq:mu-rel} in a more convenient form:

\begin{equation}
    \mu_{\textrm{rel,G}} = \mu_{\textrm{rel,H}} - (6.153, -0.092) \times (1/D_{L} - 1/D_{S}),
\end{equation}

\noindent where $D_{L}$ and $D_{S}$ are the lens and source distance, respectively, given in kpc. We use this relation in our Bayesian
analysis of the light curve (Section \ref{sec:light-curve}), with Galactic model and Keck constraints to determine the relative proper motion in the geocentric frame of $\mu_{\textrm{rel,G}} = 11.01 \pm 0.12\,$mas yr$^{-1}$. This is compared to the values determined from the light curve MCMC without the Keck constraints (e.g. the \cite{bennett:2014a} ``fast" and ``slow" solutions) of $\mu_{\textrm{rel,G}_{\textrm{fast}}} = 19.6 \pm 1.6\,$mas yr$^{-1}$ and $\mu_{\textrm{rel,G}_{\textrm{slow}}} = 11.6 \pm 0.9\,$mas yr$^{-1}$, respectively. The Keck measurement strongly favors the ``slow" solution, and we can rule out the ``fast" solution that would've given a nearby planetary-mass host with an exomoon companion.

\subsection{Gaia DR3} \label{sec:gaia_pms}
In an attempt to determine the absolute proper motion of the source and lens stars, we queried the Gaia DR3 catalog \citep{vallenari:2023a} to search for all stars within a 30\arcsec\, radius around the location of MB11262. Our initial search found 17 Gaia sources that were matched to Keck stars in both 2012 and 2021 epochs (note the MB11262 source star is not in the Gaia catalog). We further trimmed the list of Gaia sources by their Renormalized Unit Weight Error and Astrometric Excess Noise Significance (RUWE \& AENS; \cite{lindegren:2012a}). These two metrics are used as a statistical criterion to determine the general astrometric data quality in Gaia. After cutting stars with RUWE $>$ 1.4 and AENS $>$ 2 mas, there remained eight good-quality Gaia-Keck sources.

\begin{figure*}[!htb]
\includegraphics[width=0.85\linewidth]{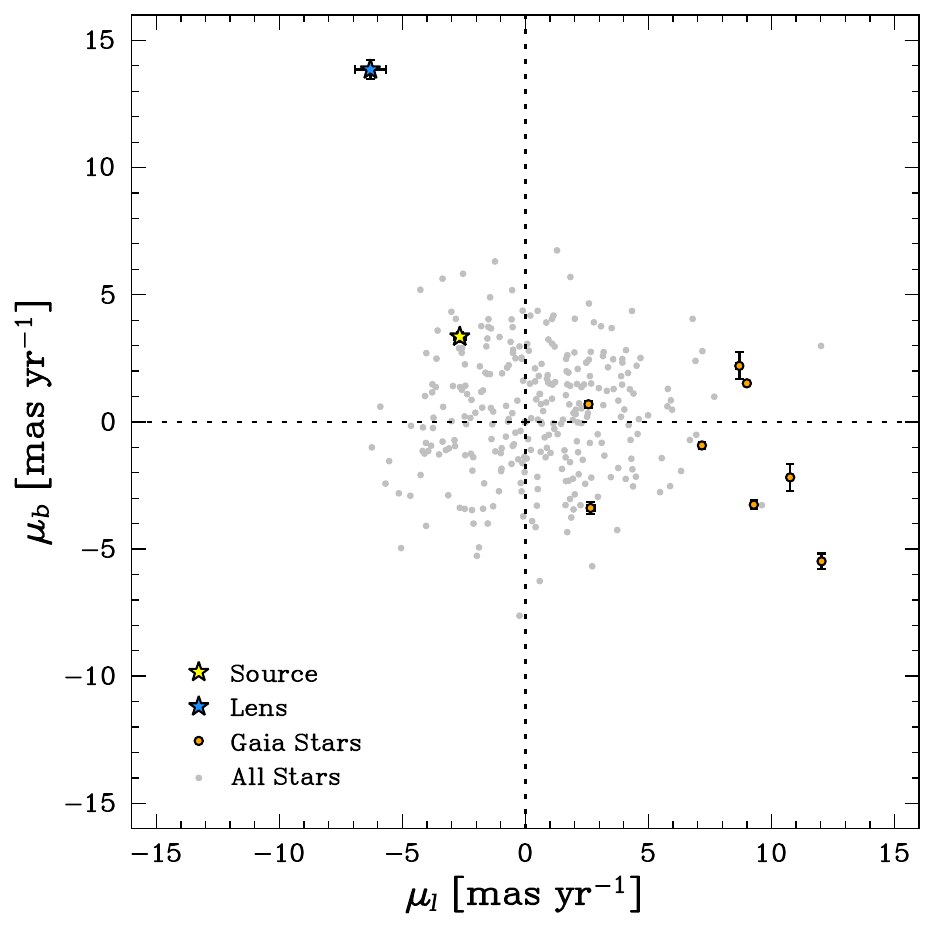}
\centering
\caption{\footnotesize Proper motions of all stars detected in both Keck epochs (grey points), given in galactic coordinates. The yellow star indicates the position of the MB11262 source, the blue star indicates the position of the MB11262 high-velocity lens. The orange points show the cross-matched Gaia-Keck reference stars used for the transformation described in Section \ref{sec:gaia_pms}. A majority of the Gaia objects are likely foreground disk stars, owing to their larger $\mu_l$ values compared to the mean motion of galactic bulge stars centered around $\mu_l \,{\sim}\, 0$ mas/yr. \label{fig:vpd}}
\end{figure*}

\subsubsection{Alignment Procedure} \label{sec:gaia_align}
The astrometric measurements extracted from the 2012 and 2021 Keck epochs must be transformed into a common reference frame in order to calculate the motion of the stars, including the source and lens. For this we follow the iterative methodology described in \cite{lu:2016a}. First, following standard image processing methods, a 2D polynomial transformation of the form

\begin{equation} \label{eq:x_affine}
    x^{\prime} = a_0 + a_1x + a_2y + a_3x^2 + a_4xy + a_5y^2 + ...
\end{equation}

\begin{equation} \label{eq:y_affine}
    y^{\prime} = b_0 + b_1x + b_2y + b_3x^2 + b_4xy + b_5y^2 + ...
\end{equation}

\noindent is applied to the images in order to match them to a reference image. Equations \ref{eq:x_affine} and \ref{eq:y_affine} are affine transformations, which are used to model translation, rotation, scaling, and shearing introduced by the different cameras (in this case NIRC2 and OSIRIS imagers). \\
\indent In the first pass, the Keck images are aligned to the absolute reference frame of Gaia with the first-order 2D polynomial transformation to establish the initial transformation. We note the Keck catalog is matched to the Gaia DR3 catalog using the pattern matching algorithm of \cite{groth:1986a}, which is described further in \cite{lu:2016a}. We subsequently perform a second pass where the Keck catalogs are aligned to themselves, using a 2D polynomial transformation going up to second order. In the second pass, the Keck catalogs are aligned to the output reference frame that was derived in the first pass, which further refines the overall reference frame and the derived proper motions.

\subsubsection{Source and Lens Absolute Proper Motion}
The precise multi-epoch Keck-Gaia alignment allows us to determine the absolute proper motions of all Keck sources measured in both 2012 and 2021 epochs, including the MB11262 source and lens stars. We determine the source proper motion in galactic coordinates:

\begin{equation} \label{eq:mu_s}
    \mu_S = (\mu_{S,l},\, \mu_{S,b}) = (-2.67 \pm 0.22, 3.35 \pm 0.19)\, \textrm{mas yr}^{-1}
\end{equation}

\noindent Following equation 1 of \cite{skowron:2014a}, this gives us the lens star proper motion:

\begin{equation}
    \mu_L = (\mu_{L,l},\, \mu_{L,b}) = (-6.30 \pm 0.62, 13.86 \pm 0.37)\, \textrm{mas yr}^{-1}
\end{equation}

\noindent Using the distance of $D_L = 7.49 \pm 0.91$ kpc (Section \ref{sec:lens-properties} and Table \ref{tab:lens_params}), we calculate a transverse velocity for the lens system of $v_L = (v_{L,l}\, , v_{L,b}) = (-224.04 \pm 21.05, 492.77 \pm 62.29)$ km s$^{-1}$, which gives a total transverse velocity of $541.31\, \pm\, 65.75$ km s$^{-1}$. The overall motion for the source star system is consistent with the bulge population of stars, however the lens system velocity is very high, particularly in the galactic latitude direction. Figure \ref{fig:vpd} shows the proper motions of all Keck stars in grey, the motion of the source in yellow and the motion of the lens in blue. The eight well-measured Gaia sources are shown as orange points. \\
\indent In the discovery paper for this event, \cite{bennett:2014a} employed a Bayesian analysis that incorporated the source star proper motion ($\mu_S$) using a method developed by \cite{skowron:2014a}. Their source proper motion estimate for MB11262, converted to the galactic coordinate system, is $\mu_{\textrm{S}} = (\mu_{S,l},\, \mu_{S,b}) = (-0.40, -2.44) \pm (2.8, 2.6)$ mas yr$^{-1}$. The longitudinal component of this motion is consistent with with the source longitudinal motion we find. However the motion in the galactic latitude direction is at contrast with what we find (eq. \ref{eq:mu_s}). Additionally, using the \cite{koshimoto:2021a} Galactic model, we calculate the prior probability of ($i$) having the lens within the 1$\sigma$ ranges of the measured $\mu_{\textrm{rel,H}}$ (Table \ref{table:epoch_seps} bottom row) and $K_L = 22.34 \pm 0.15$, and ($ii$) having an ambient star (with $K = 22.34 \pm 0.15$) within the corresponding uncertainty regions in East and North (Table \ref{table:epoch_seps} top row). In the ambient star case we multiply the prior probability by the probability of the true lens flux being fainter than the Keck limiting magnitude (e.g. $P(K_L > 22.75)\,{\sim}\,0.4007$). We find:

\begin{equation}
    \frac{P_{\textrm{amb}}}{P_{\textrm{lens}}} = \frac{9.1236\times 10^{-6}}{1.6609\times 10^{-6}} = 5.4932.
\end{equation}

\noindent According to the Galactic model, the faint signal we detect is ${\sim} 5.5\times$ more likely to be an ambient star than the lens host star. However we note the published version of the \cite{koshimoto:2021a} model does not include a population of stars from the stellar halo, which have systematically higher velocity dispersions than the disk/bulge stellar populations. We have modified the galactic model to include a stellar halo population of stars based on \cite{robin:2003a}, which did change the ambient-to-lens star relative probability from ${\sim}14\times$ to ${\sim}5.5\times$ in favor of the ambient star scenario. The \cite{koshimoto:2021a} model also lacks a population of HVSs, which is more difficult to incorporate as there have only been a few dozen confirmed HVSs and no public Galactic model includes this exceedingly small population. \\
\indent Finally, under the assumption that the detected signal is the true lens, the transverse velocity (${\sim}541$ km s$^{-1}$) is the largest detected for any planetary system (see Section \ref{sec:lens-properties} and Figure \ref{fig:vt_vs_distance}). We note that this velocity falls below the bulge population escape velocity of 600 km s$^{-1}$ \citep{brown:2008a, brown:2018a}. According to the literature, MB11262L would be below the lower limit of what may be considered an HVS, which have velocities in the range $550 - 1500$ km s$^{-1}$.  


\section{New Light Curve Modeling} \label{sec:light-curve}
\begin{figure*}
\includegraphics[width=1.0\linewidth]{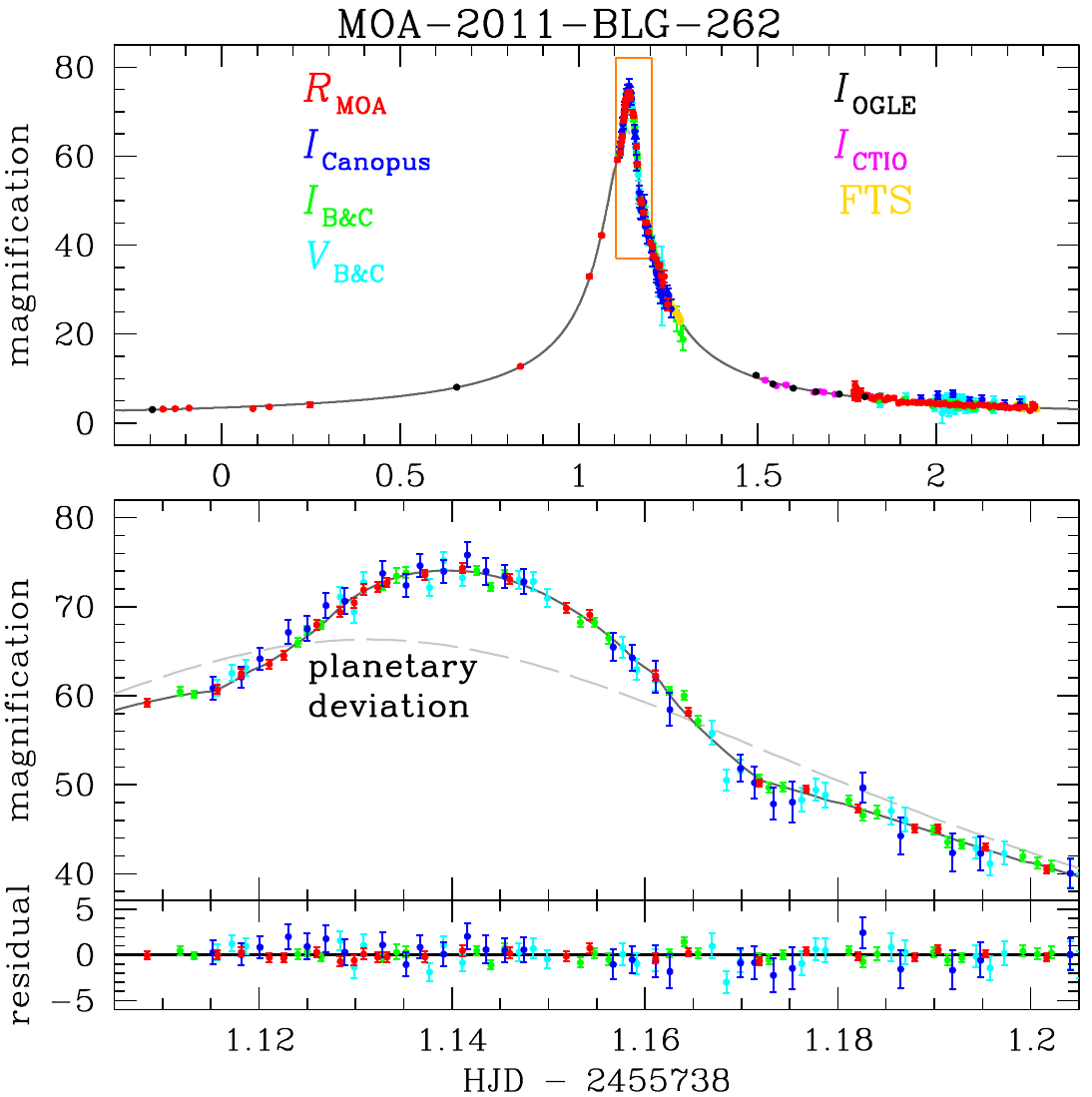}
\centering
\caption{\footnotesize Best fit planetary light curve model (solid black curve) for MOA-2011-BLG-262 with constraints from the high resolution follow-up data as described in Section \ref{sec:follow-up}. This 2L1S model is given by the best-fit values listed in column 2 of Table \ref{tab:lcpar}. A 1L1S model is shown as a dashed grey curve. \label{fig:lc}}
\end{figure*}

The light-curve modeling follows the image-centered ray shooting method of \cite{bennett:1996a} and \cite{bennett:2010a}, modified to include constraints on the brightness and separation of the lens and source stars from the high resolution imaging \citep{bennett:2024a}. This can help prevent the light curve modeling from exploring areas in the parameter space that are excluded by the high resolution follow-up observations. We refer the reader to \cite{bennett:2024a} for a full description of the methodology for applying these constraints to the light curve modeling. Figure~\ref{fig:lc} shows the light curve photometry along with the best fit two-lens one-source (2L1S) model (black curve) and Table \ref{tab:lcpar} shows the parameters of our best-fit model, as well as the MCMC averages of models 
consistent with the data. These values in Table \ref{tab:lcpar} are also compared to the results of the ``slow" solution from the original study of \cite{bennett:2014a}. 

\indent Determining the angular size of the source star ($\theta_*$) allows one to estimate the source radius crossing time ($t_*$) through the following relation:

\begin{equation} \label{eq:tstar_equation}
    t_* = \frac{\theta_*}{\mu_{\textrm{rel}}}.
\end{equation}

\noindent But first we need to determine the extinction corrected source magnitude and color. We first adopt the \cite{bennett:2014a} calibration of the CTIO $V$ and $I$ band data to the OGLE-III catalog \citep{ogle3-phot}, and locate the red clump centroid at $V_{\rm rc} - I_{\rm rc} = 15.92$, $I_{\rm rc} =2.22$, following the method of \cite{bennett:2010b}. Using the bulge red clump giant magnitude, color, and distance from \citet{nataf:2013a}, we find $I$ and $V$ band extinction of $A_I = 1.45$ and $A_V = 2.62$. Using the source magnitudes from Table~\ref{tab:lcpar}, we find extinction corrected magnitudes of $I_{S0} = 18.496$  and $V_{S0} = 19.231$. This allows us to use the surface brightness relation from the analysis of \citet{boyajian:2014a}, but we use the following custom formula \citep{aparna16} using stars spanning the range in colors that are relevant for microlensing events:

\begin{equation}
    \textrm{log}({2\theta_{*}}) = 0.5014 + 0.4197(V_{S0} - I_{S0}) - 0.2 I_{S0}
\end{equation}

\noindent This yields $\theta_{*} = 0.645 \pm 0.026\, \mu$as, which is smaller than the \cite{bennett:2014a} value of $\theta_{*} = 0.776 \pm 0.059\, \mu$as. This difference is due in part to the combination of the error in magnitude from the older DoPHOT reduction and an improved knowledge of the red clump from \cite{nataf:2013a}.

\begin{deluxetable*}{lcccc}[t]
\deluxetablecaption{Best Fit MOA-2011-BLG-262 Model Parameters\label{tab:lcpar}}
\tablecolumns{5}
\setlength{\tabcolsep}{18.5pt}
\tablewidth{\columnwidth}
\tablehead{
\colhead{Parameter} &
\colhead{Units} & \colhead{Value} & \colhead{MCMC Averages} & \colhead{\cite{bennett:2014a}}
}
\startdata
$t_E$ & days & $3.781$ & $3.788 \pm 0.125$ & $3.858 \pm 0.126$\\
$t_{0}$ & HJD$'$ & $5739.131$ & $5739.132 \pm 0.001$ & $5739.131 \pm 0.001$\\
$u_0$ & {} & $0.015$ & $0.015 \pm 0.001$ & $0.015 \pm 0.001$\\
$s$ & {} & $0.939$ & $0.939 \pm 0.007$ & $0.926 \pm 0.032$\\
$\alpha$ & radians & $1.801$ & $1.801 \pm 0.006$ & $1.811 \pm 0.006$\\
$q \times 10^{-4}$ & {} & $4.508$ & $4.499 \pm 0.025$ & $4.390 \pm 0.025$\\
$t_*{}$ & {days} & $0.022$ & $0.022 \pm 0.003$ & $0.022 \pm 0.003$\\
$I_s{}$ & {} & $19.946$ & $19.946 \pm 0.022$ & $19.937 \pm 0.041$\\
$V_s{}$ & {} & $21.851$ & $21.852 \pm 0.032$ & $21.898 \pm 0.041$\\
fit $\chi^2$ & {} & $4388.80$ & {} & $5760.85$\\
d.o.f & {} & 4386 & {} & 5199\\
\enddata
\tablenotetext{}{\footnotesize{\textbf{Notes.} HJD$'$ = HJD$-2450000$. The \cite{bennett:2014a} values are given for their best-fit close ``slow" solution.}}
\end{deluxetable*}

\indent To measure our new lens system parameters, we sum over the MCMC results using a Galactic model \citep{bennett:2014a} with weights for the microlensing rate and our $\mu_{\textrm{rel,H}}$ value from Keck (described in Section \ref{sec:prop-motion}). Additionally, we include the source distance as a fitting parameter in the re-modeling of the light curve with imaging constraints. We include a weighting from the \cite{koshimoto:2021a} Galactic model as a prior for $D_S$, and we also use the same Galactic model to obtain a prior on the lens distance for a given value of $D_S$. This prior is not used directly in the light curve modeling, but instead is used to weight the entries in a sum of Markov chain values.

\section{Lens System Properties} \label{sec:lens-properties}
\begin{figure}
\includegraphics[width=\linewidth]{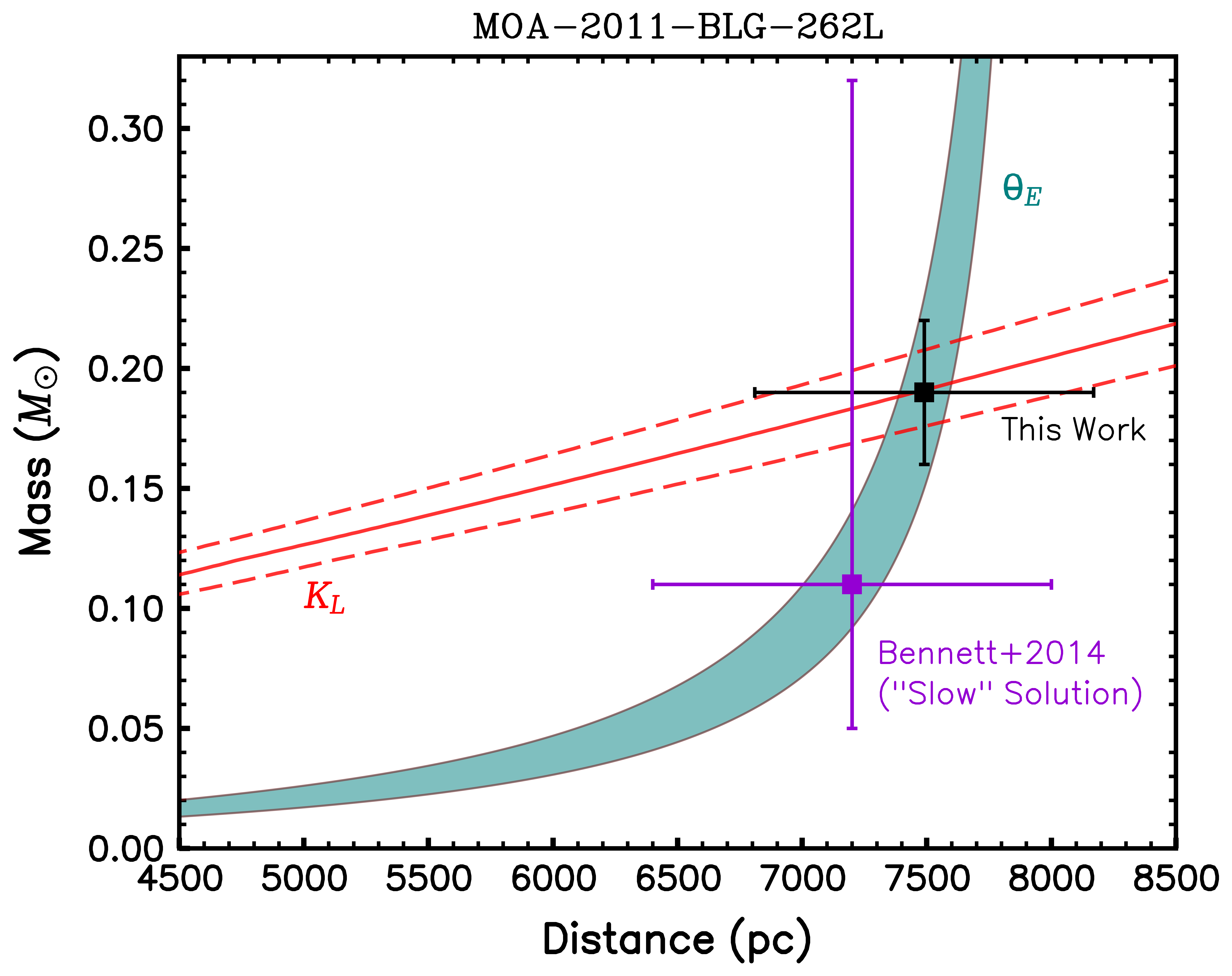}
\centering
\caption{\footnotesize The mass-distance relation for MOA-2011-BLG-262L with constraints from the lens flux measurement in Keck $K$-band (red). Dashed lines show the 1$\sigma$ error bars derived from the flux measurement uncertainty and intrinsic error in the empirical mass-luminosity relation of \cite{delfosse:2000a}. The solid teal region shows the mass-distance relation calculated using the angular Einstein radius measurement ($\theta_E$). \label{fig:MD_relation}}
\end{figure}

The angular Einstein radius, $\theta_E$ gives a relation that connects the lens system mass to the source and lens distances, $D_S$ and $D_L$ (\cite{bennett:2008b}, \cite{gaudi:2012a}). The relation is given by:

\begin{equation} \label{eq:md_thetaE}
    M_{L} = \frac{c^2}{4G}\theta_{E}^{2}\frac{D_{S}D_{L}}{D_{S}-D_{L}},
\end{equation}

\noindent where $M_{L}$ is the lens mass, $G$ and $c$ are the gravitational constant and speed of light. As mentioned previously, the measurement of $\mu_{\textrm{rel,H}}$ from the high resolution imaging allows us to measure $\mu_{\textrm{rel,G}}$ to high precision, which ultimately lets us determine $\theta_E \sim \mu_{rel,G} \times t_E$. Figure \ref{fig:MD_relation} shows the measured mass and distance for the lens planetary system (black data point). The solid red curve shows the constraint from the mass-luminosity relation derived from the lens flux measurement in the Keck Kp-band imaging. The dashed red lines show the measurement error from the Keck lens flux detection combined with the intrinsic uncertainty in the empirical mass-luminosity relation \citep{delfosse:2000a}. Additionally, the mass-distance relation obtained from the measurement of $\theta_{E}$ (i.e. Equation \ref{eq:md_thetaE}) is shown as a solid green region. Lastly, we plot the estimated mass and distance for the lens system given by the ``Slow" solution of \cite{bennett:2014a} as the purple data point in Figure \ref{fig:MD_relation}.

\begin{figure*}
\includegraphics[width=0.9\linewidth]{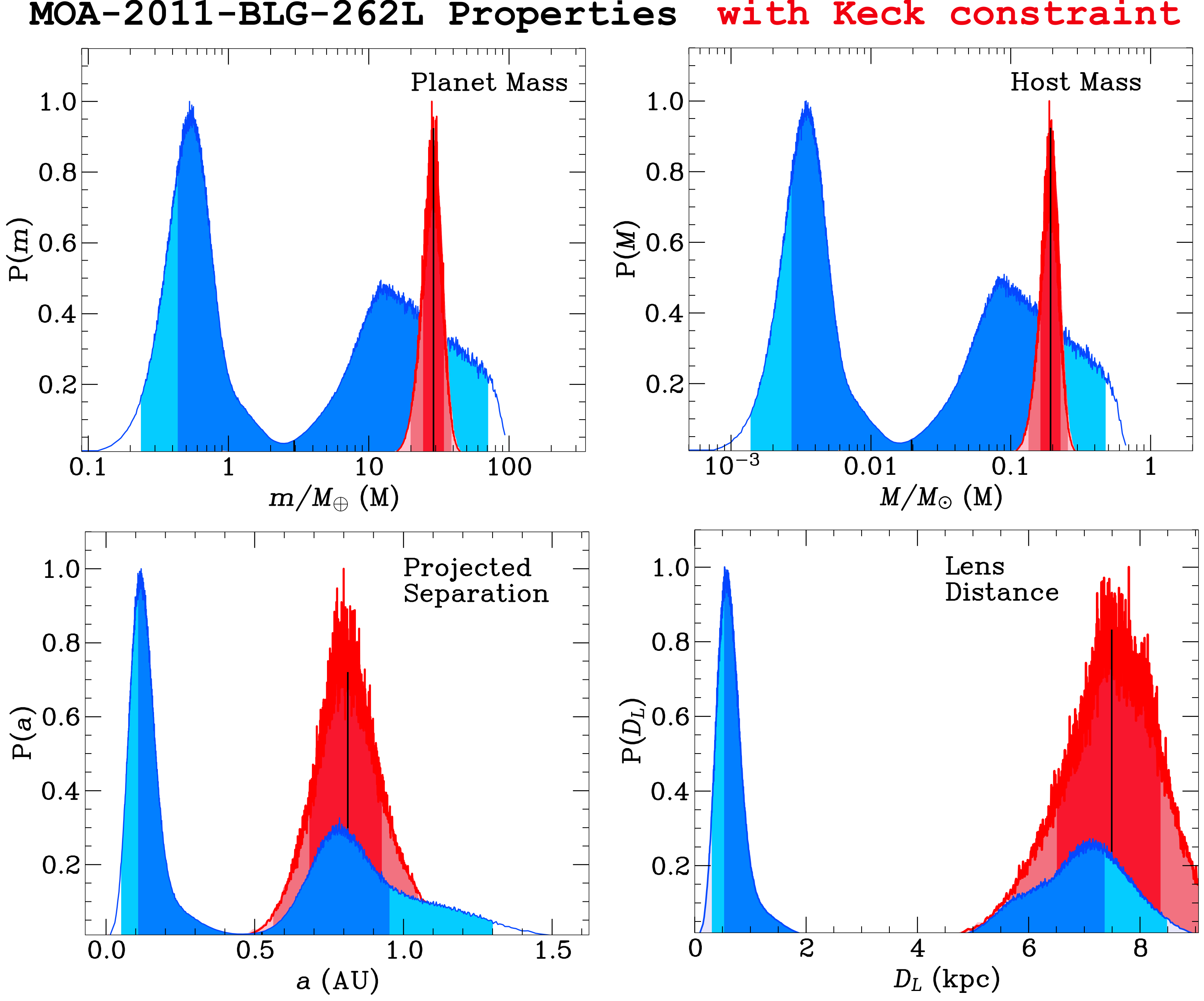}
\centering
\caption{\footnotesize Posterior probability distributions for the planetary companion mass, host mass, projected separation and the distance to the lens system are shown with only light curve constraints in blue (e.g. \cite{bennett:2014a} results) and with the additional constraints from our Keck follow-up observations in red. The central 68.3\% of the distributions are shaded in darker colors (dark red and dark blue) and the remaining central 95.4\% of the distributions are shaded in lighter colors. The vertical black lines mark the median of the probability distribution for the respective parameters. \label{fig:posteriors}}
\end{figure*}

\begin{figure*}
\includegraphics[width=0.9\linewidth]{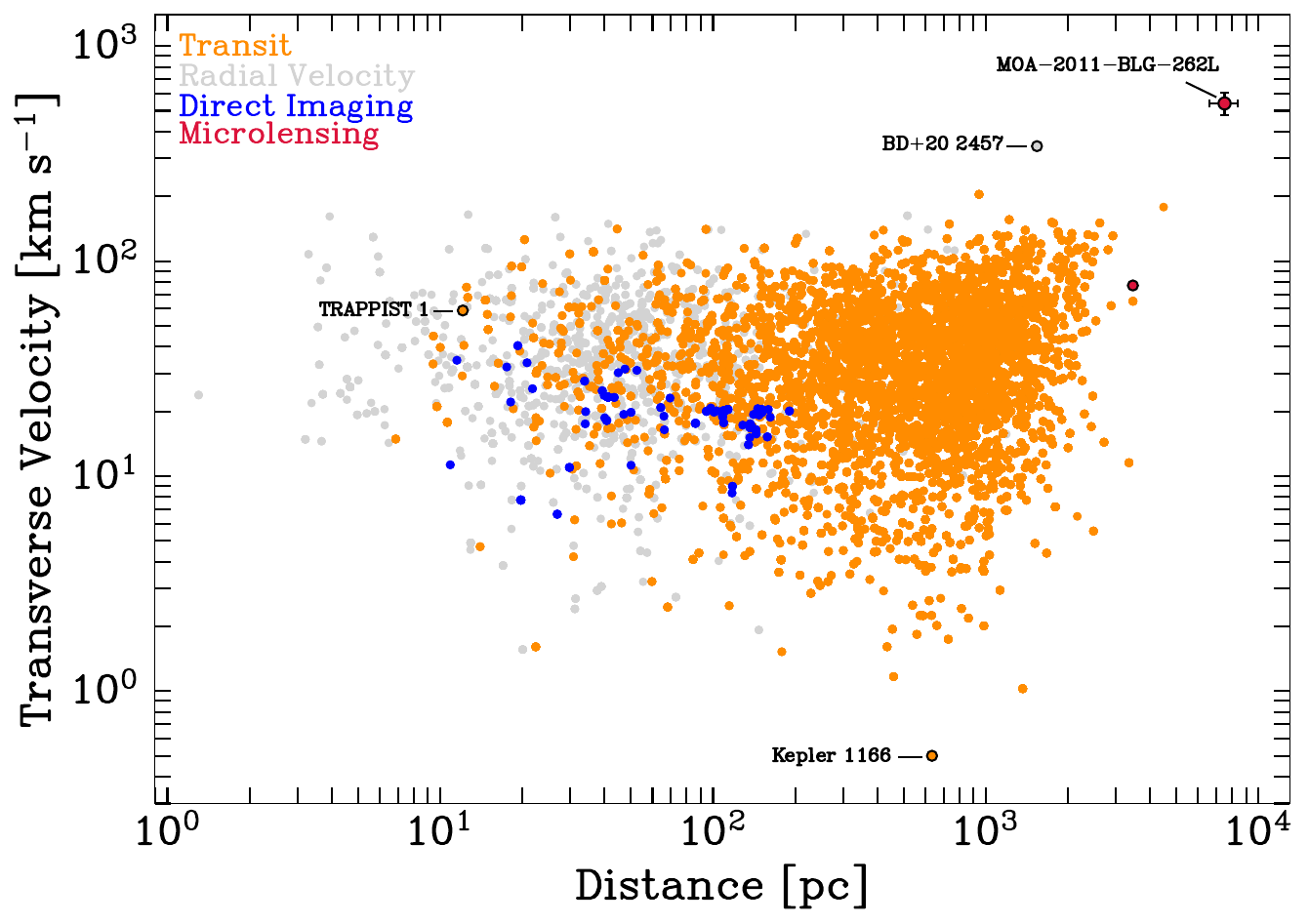}
\centering
\caption{\footnotesize Confirmed planetary systems with measured transverse velocities from The NASA Exoplanet Archive (accessed June 25, 2024). MOA-2011-BLG-262L (in red with error bars) is the highest velocity planetary system and the most distant planetary system with a measured transverse velocity. For reference, the previous highest velocity system (BD+20 2457; \cite{niedzielski:2009a}), lowest velocity system (Kepler-1166; \cite{morton:2016a}), and TRAPPIST-1 \citep{gillon:2016a} are labeled.}
\label{fig:vt_vs_distance}
\end{figure*}

\indent Table \ref{tab:lens_params} gives the derived lens system physical parameters with RMS errors and 2$\sigma$ ranges. We find that the M-dwarf lens star has a mass $M_{\textrm{host}} = 0.193 \pm 0.029M_{\Sun}$, with a sub-Saturn planetary companion of mass $m_{\textrm{planet}} = 28.917 \pm 4.753M_{\Earth}$. We can calculate the planet's semi-major axis from the expression $a_{\perp} = sD_{L}\theta_{E}$, where $s$ is the projected separation given by the light curve modeling. We find a 2D separation of $a_{\perp} = 0.813 \pm 0.121$AU. Additionally, we find the lens system is at a distance of $D_L = 7.494 \pm 0.914$ kpc, very likely located in the Galactic bulge. Figure \ref{fig:posteriors} shows the results for the physical parameters of the lens system with (red) and without (blue) the constraints from the high resolution Keck imaging. As mentioned previously, the new host mass and planetary mass results strongly favor the ``Slow" solution presented in \cite{bennett:2014a}. The blue distributions in Figure \ref{fig:posteriors} include both the slow and fast solutions weighted by Galactic model priors and the best-fit $\chi^2$ in each region of parameter space, as described in Section 6 of \cite{bennett:2014a}. \\
\indent We queried the NASA Exoplanet Archive\footnote{https://exoplanetarchive.ipac.caltech.edu/} to investigate all confirmed planetary systems with published transverse velocities. Figure \ref{fig:vt_vs_distance} shows the result of this search, shown as transverse velocity vs. distance to the systems. We include confirmed planets from transit, radial velocity, direct imaging, and microlensing searches. Note there is one other published microlensing event with a measured transverse velocity, OGLE-2005-BLG-071 \citep{bennett:2020a}. As shown in the figure, the results of our current work firmly indicate MB11262 is the highest velocity system hosting a planet, as well as the most distant planetary system with a measured transverse velocity. We stress an important point that remains is the need to obtain an additional epoch of high-resolution imaging (with either Keck, \textit{HST}, \textit{JWST}, etc) to definitively confirm the faint object we detect in the 2021 Keck data is in fact the true lens by directly measuring it's motion since the 2021 Keck epoch.

\begin{deluxetable*}{lccc}[!htp]
\deluxetablecaption{Planetary System Properties from Lens Flux Constraints\label{tab:lens_params}}
\tablecolumns{4}
\setlength{\tabcolsep}{12.5pt}
\tablewidth{\columnwidth}
\tablehead{
\colhead{\hspace{-7cm}Parameter} & \colhead{Units} &
\colhead{Values \& RMS} & \colhead{2-$\sigma$ range}
}
\startdata
Angular Einstein Radius ($\theta_E$) & mas & $0.113 \pm 0.005$ & $0.104-0.123$\\
Geocentric lens-source relative proper motion ($\mu_{\textrm{rel,G}}$) & mas/yr & $11.01 \pm 0.12$ & $10.77-11.25$\\
Host mass ($M_{\rm host}$) & $M_{\Sun}$ & $0.19 \pm 0.03$ & $0.13-0.26$\\
Planet mass ($M_{\rm p}$) & $M_{\Earth}$ &$28.92 \pm 4.75$ & $19.42-38.44$\\
2D Separation ($a_{\perp}$) & AU & $0.82 \pm 0.12$ & $0.58-1.08$\\
3D Separation ($a_{3\textrm{d}}$) & AU & $0.98^{+0.56}_{-0.20}$ & $0.64-3.89$\\
Lens Distance (D$_{L}$) & kpc & $7.49 \pm 0.91$ & $5.45-9.14$\\
Source Distance (D$_{S}$) & kpc & $7.93 \pm 0.98$ & $5.81-9.73$\\
\enddata
\end{deluxetable*}


\section{Discussion and Conclusion} \label{sec:conclusion}

\indent Our Keck AO follow-up observations have identified the MB11262L planetary host star from measurements of the host star K-band magnitude and the lens-source relative proper motion, $\mu_{\textrm{rel,H}}$. The $K_L = 22.3$ mag host star is the faintest lens directly detected in high-resolution follow-up imaging, with a lens-source flux ratio of 0.02 (e.g. the lens is $50 \times$ fainter than the source). We used the Keck measurements to constrain modeling of the light curve photometry, which allowed us to rule out the ``fast" solution of \cite{bennett:2014a} and confirm the low-mass host with a planetary companion located in the Galactic bulge. We find the host and planet have masses of $M_{\textrm{host}} = 0.19M_{\Sun}$ and $m_{\textrm{planet}} = 28.92M_{\Earth}$. 

\indent MB11262L is the lowest mass microlens host star with a confirmed mass measurement. Prior to this detection, \cite{terry:2024a} reported a direct detection of the low-mass microlens host for event MOA-2007-BLG-192 with a host mass of $0.28M_{\odot}$ orbited by a super-Earth mass planet. Another candidate for the lowest-mass microlens host star is OGLE-2011-BLG-0265 \citep{skowron:2015a} with two possible solutions of $0.14M_{\odot}$ or $0.22M_{\odot}$, although a direct lens detection for this target has not been achieved yet.\\
\indent The MB11262L system very likely resides within the Galactic bulge, and a multi-epoch analysis that includes a calibration to the Gaia absolute reference frame gives a transverse velocity of $541.3\, \pm\, 65.8$ km s$^{-1}$ for the lens star. Given the very high lens-source relative proper motion ($\mu_{\textrm{rel,H}}\, {\sim}\, 11$ mas yr$^{-1}$), an additional epoch of high-resolution data with good image quality (FWHM\,${\sim}\,60$ mas) can detect the increased lens-source separation since the 2021 measurement. This would provide a definitive confirmation of the lens and rule out scenarios where the signal is due to a source companion, lens companion, or an unrelated star. We have shown that a very faint lens host can be detected in Keck with a combination of ${\sim}$25 good-quality exposures. The \textit{RGES} survey will take approximately 40,000 exposures in the wide-Red band and 800 exposures in at least one additional passband for each field it observes during the five-year survey. This large number of dithered exposures will enable a very well-sampled PSF and yield a much higher photometric and astrometric precision compared to a few dozen Keck or \textit{HST} exposures. Thus the work presented here demonstrates that \textit{RGES} will be able to measure masses for very low mass host stars in the Galactic bulge.

\section*{Acknowledgements}
The Keck Telescope observations and data analysis were supported by a NASA Keck PI Data Award, 80NSSC18K0793, administered by the NASA Exoplanet Science Institute. Data presented herein were obtained at the W. M. Keck Observatory from telescope time allocated to the National Aeronautics and Space Administration through the agency's scientific partnership with the California Institute of Technology and the University of California. The Observatory was made possible by the generous financial support of the W. M. Keck Foundation. The authors wish to recognize and acknowledge the very significant cultural role and reverence that the summit of Maunakea has always had within the indigenous Hawaiian community. We are most fortunate to have the opportunity to conduct observations from this mountain. This work was supported by the University of Tasmania through the UTAS Foundation and the endowed Warren Chair in Astronomy and the ANR COLD-WORLDS (ANR-18-CE31-0002). M.J.H. acknowledges support from the National Science Foundation under grant No. 1909641 and the Heising-Simons Foundation under grant No. 2022-3542.


\textit{Software}: Astropy \citep{robitaille:2013a}, DAOPHOT-II \citep{stetson:1987a}, daophot$\_$mcmc \citep{terry:2021a}, eesunhong \citep{bennett:1996a}, genulens \citep{koshimoto:code}, KAI \citep{lu:2022a}, Matplotlib \citep{hunter:2007a}, Numpy \citep{oliphant:2006a}, SWarp \citep{bertin:2010a}.

\bibliographystyle{aasjournal}
\bibliography{Terry_mb11262.bib}

\end{document}